\documentclass[twocolumn,aps]{revtex4}
\usepackage[T1]{fontenc} 

\usepackage{color}
\usepackage{ulem}
\usepackage{graphicx}

\newcommand{\rd}{{\rm d}}
\newcommand{\ri}{{\rm i}}

\begin{document}

\author{F. Geesmann}
\author{P. Thurau}
\author{S. Rodehutskors}
\author{T. Ziehm}
\author{L. Worbes}
\author{S.-A. Biehs}
\email{s.age.biehs@uol.de}
\author{A. Kittel}
\email{kittel@uol.de}
\affiliation{Institute of Physics, School of Mathematics and Science, Carl-von-Ossietzky University Oldenburg, Oldenburg}

\title[Transition from near-field to extreme near-field radiative heat transfer]
  {Transition from near-field to extreme near-field radiative heat transfer}


\begin{abstract}
The radiative heat transfer in the extreme near-field regime, i.e.\ for distances
below 10~nm, remains poorly understood. There are competing experimental results in this regime, with some in good agreement with theoretical predictions, while others report drastically elevated heat fluxes, orders of magnitude larger than what theory suggests. Whether the theory of fluctuational electrodynamics can predict the radiative heat transfer in this extreme near-field regime or not remains a matter of active debate. In this study, a radiative heat transfer measurement is presented between a gold-coated sphere and a gold film sample in the transition regime between the near-field and the extreme near-field regime just before contact. The radiative heat flux measurement is made by using a near-field scanning thermal microscope equipped with a temperature sensor at a sharp tip as a heat flux sensor. 
We find an excellent agreement with the theoretical predictions of fluctuational electrodynamics in the near-field regime. In the extreme near-field regime however, a highly increased radiative heat flux is observed with values about 100 times larger than the theoretical predictions, indicating that fluctuational electrodynamics fails 
to capture the radiative heat flux in this regime. 
\end{abstract}


\maketitle

It is a fact that for distances smaller than the thermal wavelength $\lambda_{\rm th}$, which is about 10 microns at room temperature, radiative heat transfer between two bodies is expected to deviate from the blackbody limit. This led to the long-standing prediction that for near-field distances, the amount of heat exchanged between two real materials can surpass the Stefan-Boltzmann value by orders of magnitude~\cite{PvH,Joulain2,Volokitin}. Over the past few decades, this prediction has been confirmed by numerous experimental setups, which can be grouped in plate-plate~\cite{HuEtAl2008,OttensEtAl2011,Kralik2012} and sphere-plate~\cite{NanolettArvind,ShenEtAl2008,NatureEmmanuel} measurements. The observed increase of the exchanged radiative heat flux in the near-field regime can be explained by the contribution of so-called evanescent waves, which are either frustrated total internal reflection modes or surface modes, i.e., surface phonons and plasmon polaritons~\cite{Volokitin,SABGreffet}. These contributions are not accounted for in the classical Planckian blackbody theory. The experimental plate-plate setups have been improved~\cite{ShenEtAl2012,Gelais,WatjenEtAl2016,BernadiEtAl2016,Song,Gotsmann,Ghashami,Lim,Rincon}, so that the theoretical predictions could be tested down to distances of about a few tens of nanometers. These experiments observe a good agreement with calculations predicting heat fluxes, which are more than 1000 times larger than the blackbody value~\cite{Fiorino2018,Salihoglu2020,Rincon}.

However, in the extreme near-field regime corresponding to distances much smaller than 10~nm, there are still many open questions concerning both the theoretical predictions of the heat transfer across a vacuum gap, as well as the experimental results. This is due to the fact that the theoretical predictions are based on Rytov's theory of macroscopic fluctuational electrodynamics~\cite{PvH,Joulain2,Volokitin}, which is expected to fail for microscopic distances because it does not describe the crossover from near field heat radiation to heat conduction.  Several recent theoretical works have explored this crossover regime by including effects like tunneling of acoustic phonons~\cite{PrunnilaEtAl2010, BudaevBogy2011, SellanEtAl2012,ChiloyanEtAl2015,PendrySasihithlu} and quantum effects due to the overlap of the electronic wave functions~\cite{XiongEtAl2014}. They showed that the radiative heat flux can be further enhanced by several orders of magnitude at distances of a few nanometers or even on the sub-nanometer scale. So far, only a single indirect measurement by Altfeder {\itshape et al.}~\cite{AltfederEtAl2010} has claimed to measure phonon tunneling. Another experiment with a near-field scanning thermal microscope (NSThM)~\cite{Kloppstech2017} has directly measured the heat flux in the extreme near-field regime ranging from 0.5~nm to about 7~nm. The results indicate a radiative heat flux which is approximately 1000 times greater than the predictions based on Rytov's theory of macroscopic fluctuational electrodynamics~\cite{PvH}. These findings have renewed the interest in the theoretical discussion of phonon tunneling or heat transfer in the extreme near-field using analytical models~\cite{VolokitinMetalsExtreme,VolokitinAccousticWaves,VolokitinDoubleLayer,Geng2022,Geng2023} and numerical methods~\cite{AlkurdiaEtAl2020,GuoEtAl2022,LiEtAl2023}. Moreover, combinations of analytical, semi-analytical and molecular dynamics results are used to discuss not only the impact of phonon tunneling, but also the possible heat flux due to electron tunneling~\cite{MessinaEtAl2018,TogunagaEtAl2021,Mauricio2023,MauricioEtAl2023b}. Even though some heuristic approaches~\cite{Henkel} can model the experimental results of Kloppstech {\itshape et al.}~\cite{Kloppstech2017} in the full range of 0.5~nm to 10~nm, other models only find a heat flux enhancement due to phonon or electron tunneling for distances smaller than about 1~nm. These discrepancies have raised questions about the validity of the measurements in Kloppstech {\itshape et al.}~\cite{Kloppstech2017} and the capability of the therein used NSThM to correctly measure heat fluxes. These doubts were further increased by another experiment reporting radiative heat fluxes down to 0.2~nm distance~\cite{Cui} in good agreement with fluctuational electrodynamics.
There is a recent experiment which also finds elevated radiative heat fluxes in the extreme near-field~\cite{JarzembskiEtAl2022} supporting rather the observations of Kloppstech {\itshape et al.}~\cite{Kloppstech2017} that fluctuational electrodynamics is not sufficient to describe heat transfer in the extreme near-field regime.

\begin{figure}
	\includegraphics[height=0.45\linewidth]{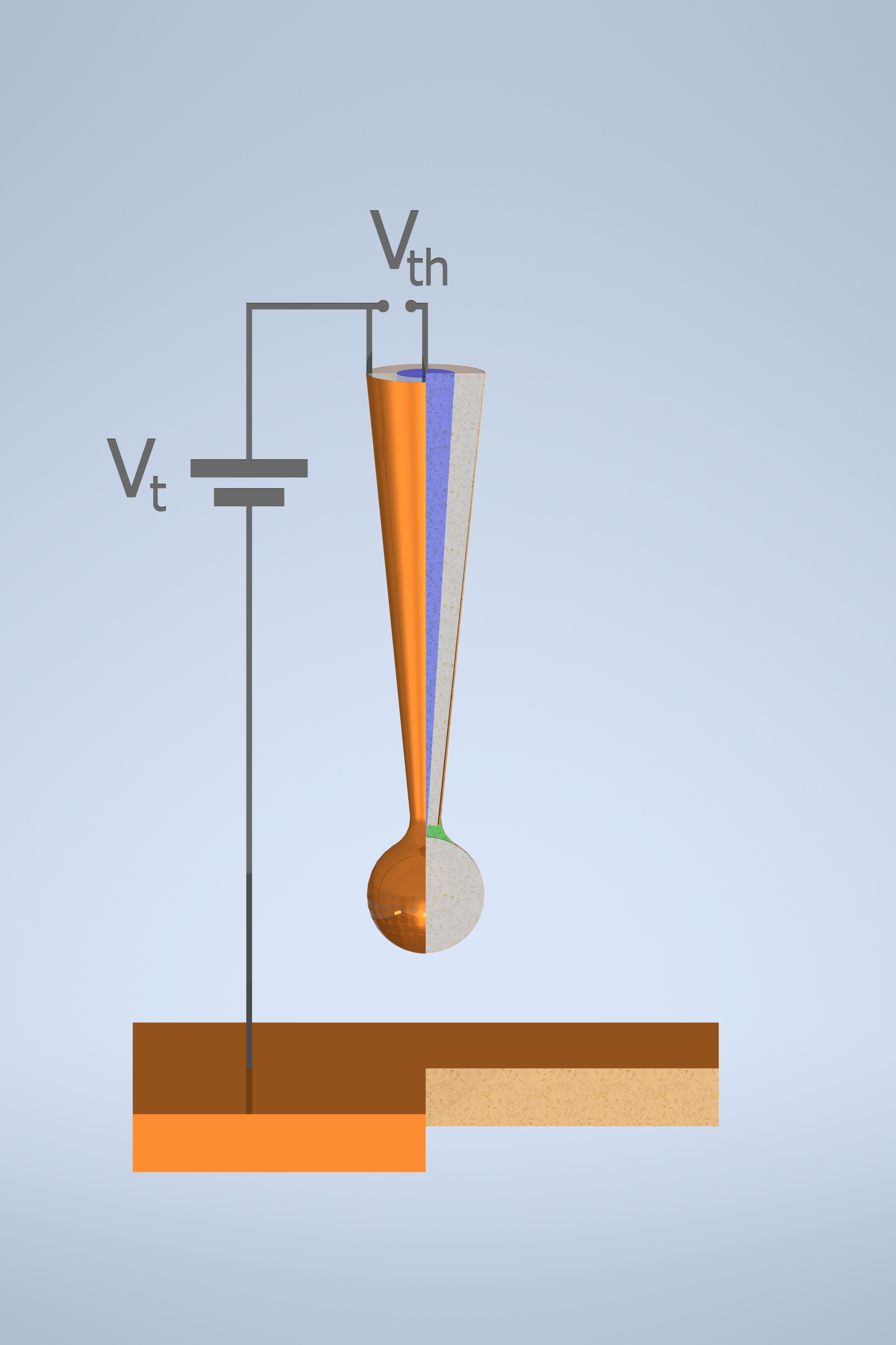}
	\includegraphics[height=0.45\linewidth]{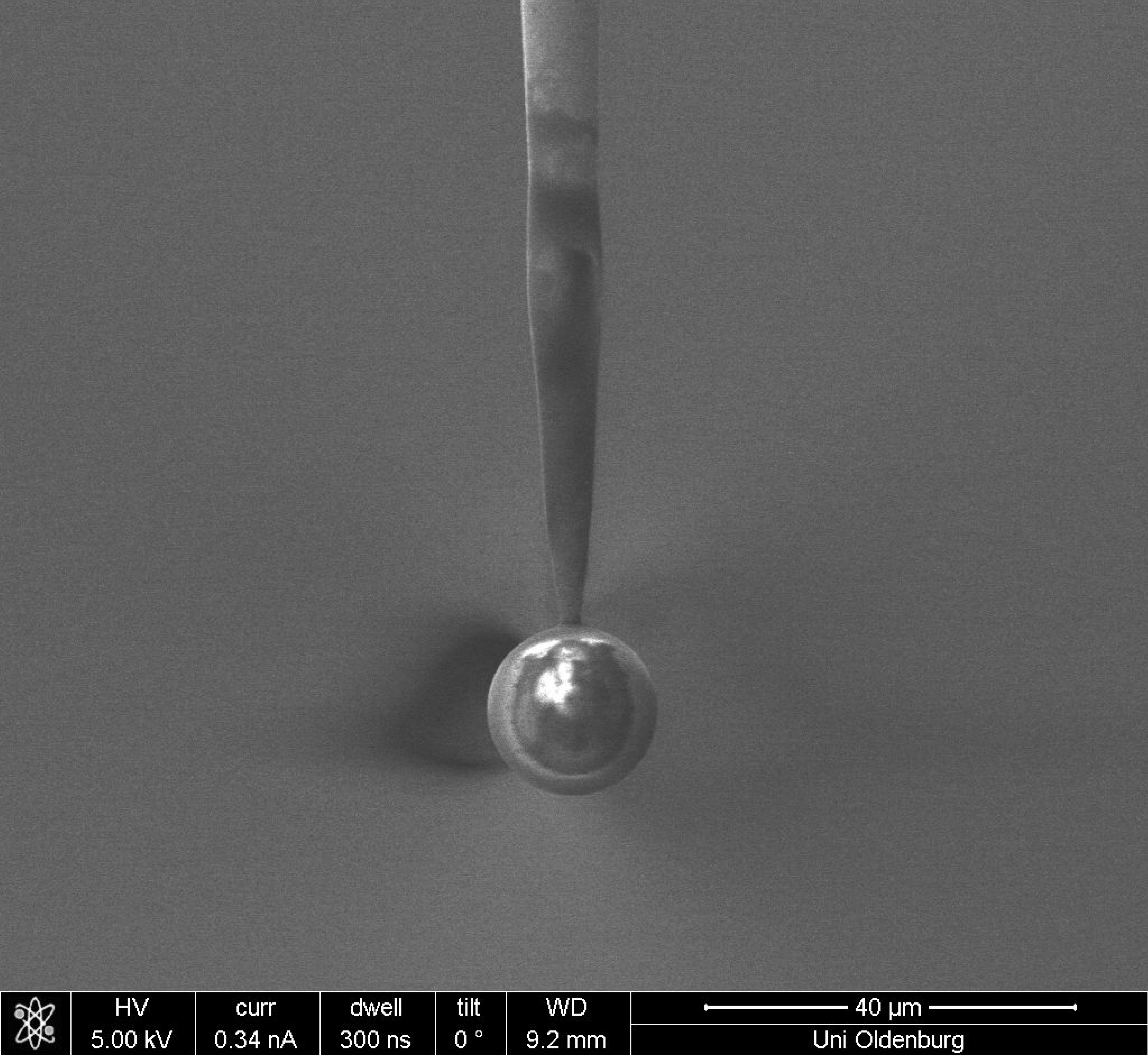}
	\caption{Left: Sketch of the setup utilizing the NSThM with an attached gold-coated sphere to apply the tunneling voltage $V_{\rm t}$ and to measure the thermoelectric voltage $V_{\rm th}$. Right: SEM image of the NSThM {probe} with attached gold-coated silica sphere.}
	\label{fgr:sketch}
\end{figure}

In this work a precise experimental observation {is presented covering} the near-field radiation and the transition between the near-field and the extreme near-field at distances of a few nanometers. The experiment was carried out with a NSThM as {it was} used in Kloppstech {\itshape et al.}~\cite{Kloppstech2017}. In that experiment heat fluxes between a sharp gold coated NSThM tip with typical tip radii of about 30~nm and a gold surface had been measured. One advantage of this small tip radius is the high lateral resolution of around 5~nm~\cite{Worbes2013}, but it also restricts the heat flux measurement between the NSThM tip and the sample surface to distances of about 7~nm. Even though the sensitivity of the NSThM with a reported value of 24~pW~K$^{-1}$ at 50~Hz bandwidth is quite high~\cite{Kloppstech2017}, it remains insufficient to measure any signal for distances larger than 7~nm, restricting its range of operation to the extreme near-field regime. This limitation of the NSThM to measure in a distance regime where fluctuational electrodynamics should certainly work has also produced some doubts on the principle capability of the NSThM to correctly measure near-field heat fluxes. Therefore, to demonstrate the transition from the near-field to the extreme near-field regime, a gold coated silica sphere with a diameter of 20~microns was attached to the NSThM sensor tip as illustrated in Fig.~\ref{fgr:sketch}. Therewith, the overall near-field heat flux between the NSThM and the sample is increased and the detectability limit due to a limiting signal to noise ratio is shifted to a larger distance. This allows measurements of heat fluxes up to distances as large as 250~nm. Furthermore, using the STM capability of the NSThM, a tunneling voltage can be applied between the gold-coated silica sphere and the gold sample, enabling precise determination of the zero distance by detecting the tunneling current across the gap between the foremost atom of the gold coated sphere and the gold surface. Details on the preparation of the coaxial-thermocouple temperature sensors with the sphere can be found in the End Matter (EM).

\begin{figure}
	\includegraphics[width=0.95\linewidth]{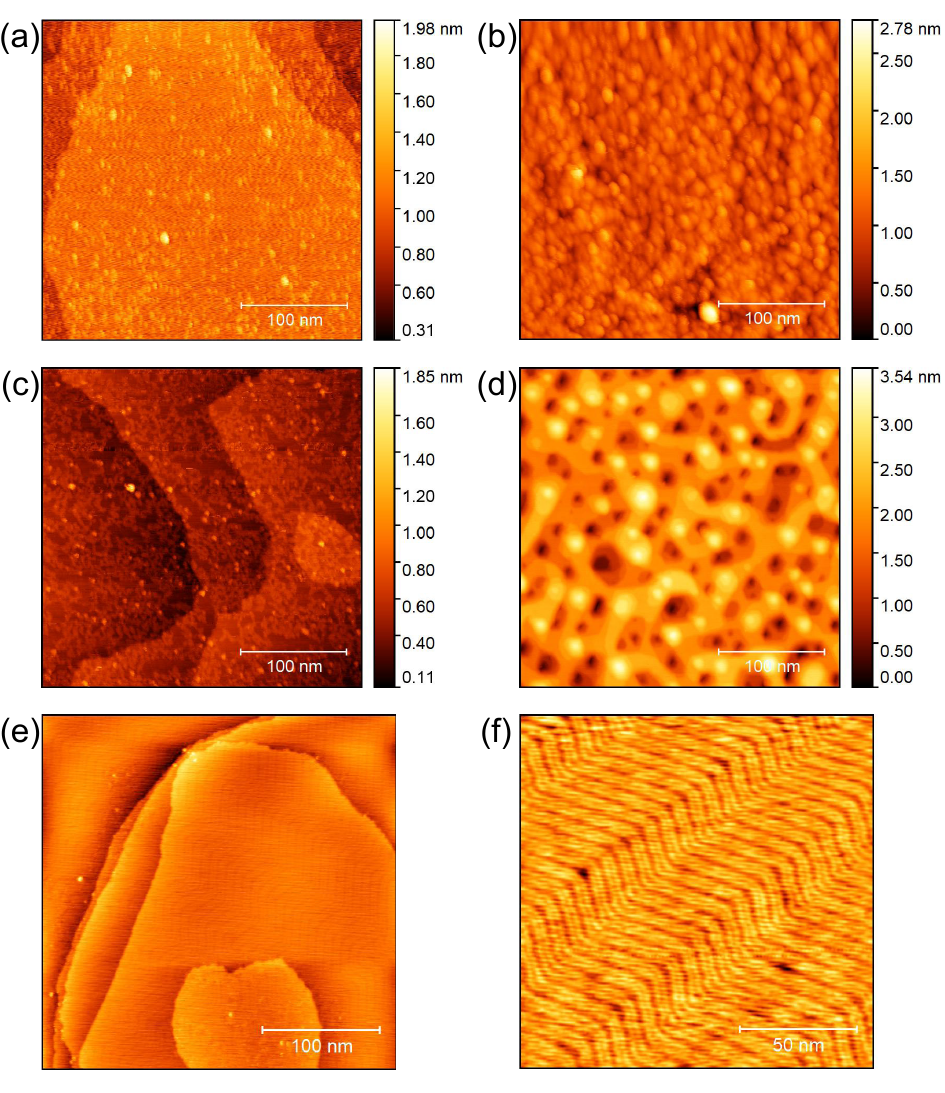}
	\caption{{STM images of the gold sample after the different stages of cleaning, scanned
			with a Tungsten tip at a tunneling setpoint of 1 nA and a tunneling bias of 800 mV. Gold
			after (a) first annealing in HV and UHV, (b) first sputtering of the sample, (c) second
			UHV annealing, (d) second sputtering of the sample, (e) third and final annealing.
			(f) Magnified view of herringbone surface reconstruction.}
	}
	\label{Fig:cleaning_effects}
\end{figure}

In order to carry out precise and reliable measurements, a clean and atomically flat sample surface is essential. For this reason, the effect of different cleaning steps on the surface will be documented in this paragraph. After loading the probes or samples into the vacuum system, they are initially annealed in high vacuum to remove the largest contaminations, as indicated by high outgassing rates, before transferring them to UHV (for details, see SI). A second annealing step is taken in UHV in order to obtain large atomically flat regions on the sample surface, as shown in Fig.~\ref{Fig:cleaning_effects}(a). The small bright spots indicate accumulations of remaining contaminants. To remove them, the sample is transferred from UHV to HV to perform in-situ argon ion sputtering, the result of which has been imaged in Fig.~\ref{Fig:cleaning_effects}(b). The rough surface morphology is a direct result of the ion bombardment during sputtering. To smooth out the sample surface, it is again annealed in UHV and scanned (cf. Fig.~\ref{Fig:cleaning_effects}(c)). These steps are repeated as required until a clean and smooth surface is produced on which herringbone surface reconstruction can be seen without contaminant surface atoms (cf. Fig.~\ref{Fig:cleaning_effects}(e) and (f)). The sample is then ready for thermal radiation measurements for several days, after which a recontamination of the surface can be observed which requires a repetition of the cleaning procedure (cf. Supplemental Material (SM)).

\begin{figure}
	\includegraphics[width=0.95
	\linewidth]{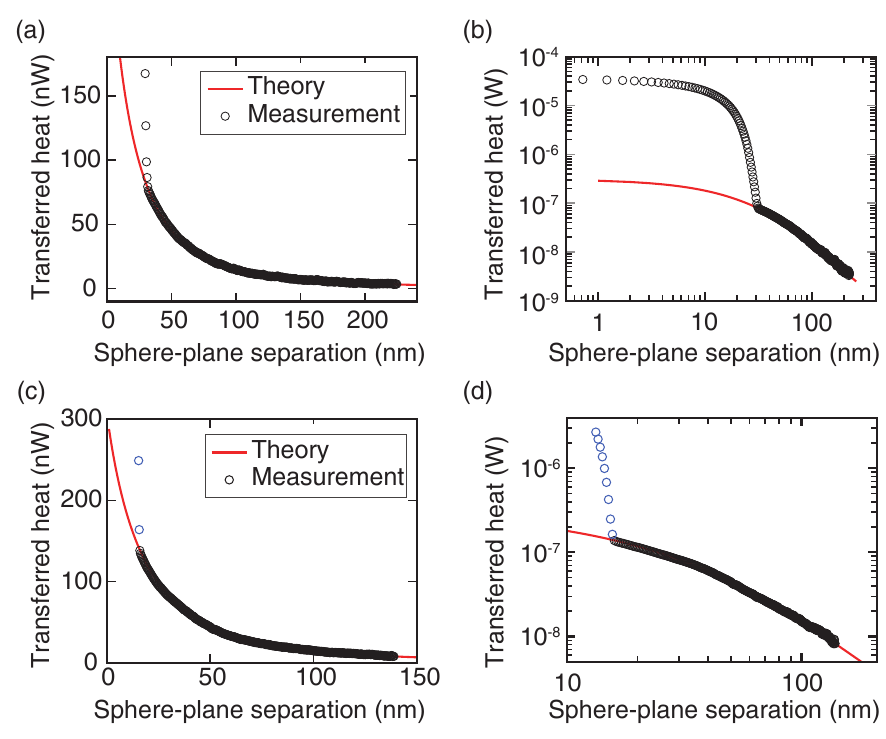}
	\caption{Experimental results of near-field heat flux compared to theoretical prediction of fluctuational electrodynamics: results of measurement without cleaning the NSThM tip and attached sphere on a (a) linear and (b) logarithmic scale;  results of measurement after cleaning the NSThM tip by means of argon-ion sputtering on a (c) linear and (d) logarithmic scale (points for which a tunneling current was observed are marked in blue).}
	\label{Fig:Data}
\end{figure}

The distance dependence of the heat transfer was measured in an ultra high vacuum STM (Omicron VT-50) with a modified scanner unit. In this setup, the atomically flat gold sample was cooled down to about 150~K by a liquid nitrogen flow cryostate. The operating pressure during the measurements was $2.0\times10^{-10}$~hPa. The tip, equipped with the sphere, was positioned above a gold surface, which was cleaned by {\itshape in-situ} argon-ion sputtering and thermal annealing. In the following, the tip was gradually moved towards the surface until a set-point of the tunneling current could be observed. During this approach, the thermoelectric voltage was acquired at each distance step. Once contact was established, defined by the tunneling current set-point of 1~nA, the tip was retracted by a preset distance before initiating the next approach. The first few approach curves with the sphere attached to the tip were measured right after fabrication without any cleaning procedure. Afterwards, the tip was cleaned by {\itshape in-situ} argon-ion sputtering and, thereafter, further approach curves were recorded. The obtained results are presented in Fig.~\ref{Fig:Data}. 

The theoretical results (details of the theory cf. EM) are compared to the experimental results. For this comparison, the expression
\begin{equation}
	P_{\rm ex}(z + z_0) = \frac{V_{\rm th}(z) + V_0}{s}
	\label{Eq:Ex}
\end{equation}
is used to fit the experimental data. The fit parameter $z_0$ has values of $37 \pm 19\,{\rm nm}$, which are much smaller than those observed by Rousseau~\cite{NatureEmmanuel}. This value becomes even smaller, $z_0 = 18 \pm 6\,{\rm nm}$, when cleaning the gold sphere by {\itshape in-situ} argon-ion sputtering. The fit parameter $s$ has a value of $s = 2.4 \pm 0.003 \,{\rm V/W}$ for all 23 measurements, taken on different days. A previous, precise calibration measurement for the sharp tips in the NSThM under realistic experimental conditions yield values of $s$ for more than twenty individual sharp tips in the range of $s = 2.4 \pm 0.19\,{\rm V/W}$\cite{Kloppstech2015}. Consequently, the fitted parameter $s$ from the data presented here is in excellent agreement with the former calibration measurements. This result is expected, as the sphere attached to the foremost end of tip should not affect the conversion factor. The heat flux must pass through the same sensor elements, producing the same thermoelectric voltage.

The experimental data is shown in Fig.~\ref{Fig:Data}(a) with a double-linear and (b) with a double-logarithmic scale in comparison to the theoretical prediction from fluctuational electrodynamics. At larger separations, above about 40~nm, the theory matches the experimental observations perfectly. When entering the extreme near-field regime, however, an increased heat flux is observed, which could be due to some contamination, because no tunneling current is detected. It should be emphasized that a mechanical gold-gold contact can be excluded because the tunneling current is practically non-zero for distances smaller than 1~nm only. Fig.~\ref{Fig:Data} shows the results of cleaning the gold sphere by {\itshape in-situ} argon-ion sputtering, in (c) with double-linear and (d) with double-logarithmic scale. We observe an excellent agreement with theory for distances larger than about 18~nm and a rapid increase in the heat flux signal for distances below 18~nm. Hence, even after the argon-ion cleaning, there is still an increased heat flux in the extreme near-field regime up to 100 times larger than the prediction by fluctuational electrodynamics, as clearly shown in Fig.~\ref{Fig:Data}. 

\begin{figure}
	\includegraphics[width=0.95
	\linewidth]{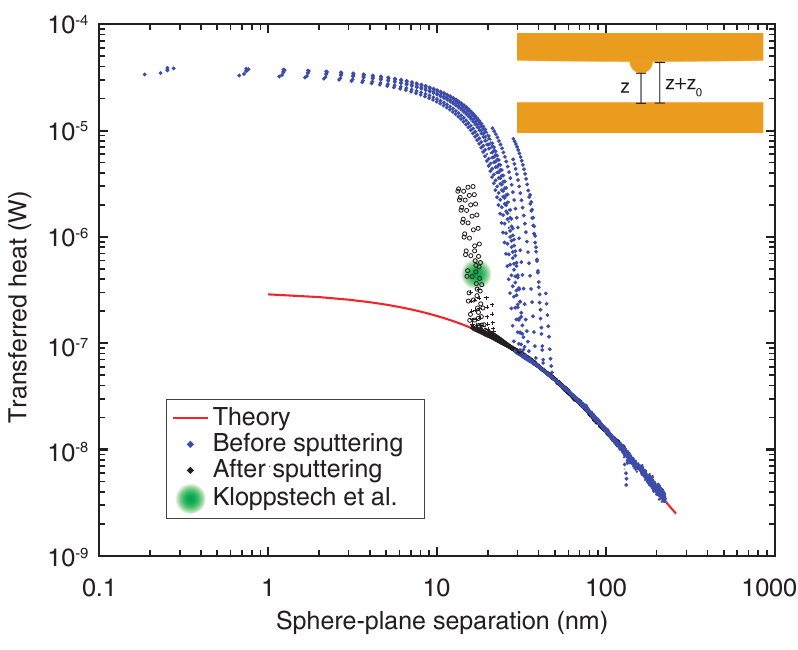}
	\caption{Comparison of the transition towards extreme near-field with and without cleaning the NSThM tip and attached sphere. The graph contains data of 23 approach curves taken on different days before (blue) and after (black) {\itshape in-situ} argon sputtering. For the measurement with a cleaned sample and a cleaned sphere, the traces with open symbols and crosses correspond to different directions of the applied tunneling voltage bias. The green disk marks the values of the transferred heat measured by Kloppstech~{\itshape et al.}~\cite{Kloppstech2017} in the range of the tunneling current for a sharp sensor tip, i.e.\ without a sphere. {For comparison, $z_0=18$~nm was added to the original $z$ values of Kloppstech~{\itshape et al.}, i.e.\ they are shifted along the abscissa by 18~nm. Inset: visualization of the described sphere-plane separations for an ideal sphere $z+z_0$ and a sphere with a protrusion, which causes the detected tunneling current at separations $z<1$~nm.}}
	\label{Fig:Data_all}
\end{figure}

To provide an overview, all acquired {data points of the recorded} curves are plotted together in a single diagram (cf. Fig.~\ref{Fig:Data_all})  {without any form of averaging or smoothing}. The curves before (marked in blue) and after argon-ion sputtering (marked in black) are easily distinguishable. While in case of the uncleaned sensor the heat transfer increases at larger gap sizes, the increase is more drastic and occurs at smaller gap sizes for the cleaned sphere. Furthermore, a tunneling current was only observed during the steep increase in heat transfer for the argon-ion sputtered sphere. To crosscheck the influence of hot or cold electrons during the tunneling process, the polarity of the tunneling voltage has been reversed. The measurements (cf.~Fig.~\ref{Fig:Data_all} marked in black) of the forward polarity are marked with open circles while the backward polarity is marked by crosses. The polarity of the tunneling voltage has no visible influence on the heat transfer for these experiments. For comparison, a green disk in  Fig.~\ref{Fig:Data_all} marks the values of the transferred heat measured by Kloppstech~{\itshape et al.}~\cite{Kloppstech2017} in the range of the tunneling current for a sharp sensor tip, i.e., without a sphere. {In order to compare the different experimental results, the thermal powers observed by Kloppstech~{\itshape et al.} were plotted with an added offset of $z_0=18$~nm, determined by the measurements of the sputtered sphere from above, thus $P_{\rm th}(z+z_0)$ is shown in the graph.} The observed heat transfer values match well, even though the exact geometry of the protrusion in the gap of the sphere on the nanometer scale is unknown. Therefore, it is impossible to specify the size of the actual interaction area in comparison to the one of Kloppstech~{\itshape et al.}~\cite{Kloppstech2017}. 

The following section discusses the implausibility of heat conduction as the reason for the short range heat transfer increase. Firstly, let it be assumed that the heat conduction at short distances is caused through a direct gold-gold contact. The heat conduction in gold is dominated by the electrons and determined by the thermal conductance quantum per channel:\cite{Pendry1983}
\begin{equation}
	G_{\rm Q} = \frac{\pi k_{\rm B}^2T}{3\hbar}.        
		\label{Eq:HQ}
\end{equation}
This means that more than 30 heat-conduction quantum channels in parallel would be required to observe a heat flow of the order of 1~\textmu W at a sample temperature $T_{\rm s}=150$~K, with a tip temperature $T_{\rm t}=298$~K resulting in a temperature difference $\Delta T=148$~K and an averaged temperature of 224~K. Consequently, with the conductance quantum $G_0=2e^2/h$ and an applied tunneling voltage of $V_{\rm t}=200$~mV, this would result in an electrical current of more than 15~\textmu A. The actually observed current was always less than 50~nA, i.e., more than 300 times smaller than what would be needed for heat conduction through a gold-gold contact able to explain the magnitude of the observed heat transfer. 

Secondly, one may assume that the heat transfer is caused by heat conduction through molecules on the sphere or sample surface. One would estimate that about 230 molecules in parallel, each contributing with about 30~pW/K~\cite{Mosso2019, Cui2019}, are necessary to reach a total power of 1~\textmu W. This might be possible in the situation before argon-ion sputtering (cf. Fig.~\ref{Fig:Data}(a) and (b)). Here, a maximal power of 30~\textmu W is measured at the smallest separation without observing any current ($I_t<100$~pA). This corresponds to about 7000 absolutely nonconducting molecules, i.e., the conductivity of each molecule has to be less than $10^{-13}$~S. Tunneling through a single layer of molecules has to be excluded, too. The molecules are too short to bridge the gap between the sphere and the sample surface, which is larger than 10~nm (cf. the fitted values of $z_0$). This means, the molecules are only able to bridge the gap between the protrusion and the sample surface. The protrusion would need to be more sharply curved than the sphere itself, leading to a much smaller surface area in the vicinity of the sample surface, whose separation has to be smaller than the length of a molecule. Furthermore, the surface of the protrusion must be covered by molecules that are all reaching to the sample surface. The molecules have to be squeezed into the surface of the sample while further approaching without resulting in an electrical or even tunneling contact. Although this sounds implausible, it can not be excluded in the situation without cleaning, but after {\itshape in-situ} argon-ion sputtering, this is no longer the case.

To summarize, measurements of the near-field radiative heat transfer were reported between a gold-coated silica sphere and a gold sample in the transition regime between the near-field and the extreme near-field {using a precisely calibrated sensor in UHV}. It was observed that for distances larger than 30~nm, the theory and experiment are in excellent agreement. However, a strong increase of the heat flux in the extreme near-field for distances below 18~nm was observed, which is in agreement with previous NSThM~\cite{Kloppstech2017} and recent shear force microscopy measurements~\cite{JarzembskiEtAl2022}. This is in contrast to recent reports of experimental results of {Kim {\itshape et al.}~\cite{Kim2015} and Cui {\itshape et al.}~\cite{Cui}} where such an increased signal vanished after tip cleaning by means of either controlled collision or {\itshape ex-situ} oxygen plasma treatment. They reported no contradiction to fluctuational electrodynamics down to 0.2~nm after tip cleaning measuring a mean conductance of 0.5~nWK. This conflicts with theoretical models for phonon and electron tunneling, which predict a steep heat flux increase for distances below 1~nm.\cite{Tokunaga2021} It was demonstrated that the strong heat flux increase persists even after cleaning by {\itshape in-situ} argon-ion sputtering. The difference in the observed results before and after cleaning emphasizes the importance of {\itshape in-situ} cleaning procedures for measurements at separations of a few nanometers. {The studies presented here document the transition between near-field and extreme near-field, but are not able to elucidate the details of the extreme near-field effect. This is due to the fact that the detailed geometry of the protrusion is unknown, which is taking over the dominating role in heat transfer at small separations. To this end, it is necessary to obtain the results from sharp tip experiments.} One can state all experiments in the extreme near-field regime are in one way or another in disagreement with existing theoretical models: The present findings and those in Refs.~\cite{Kloppstech2017,JarzembskiEtAl2022} predict an enhanced near-field heat transfer in the extreme near-field regime, which cannot be quantitatively described by existing models for phonon or electron tunneling. The experimental results presented here suggest that the heat flux in the extreme near-field is still not well understood and more systematic studies are necessary before a distinct experimental observation of the heat flux contribution of phonon or electron tunneling can be made.



\section*{Acknowledgement}
This work has received founding from the European Community through the Horizon 2020 research 
and innovation programs under grant agreement No. 766853 (EFINED). The authors further acknowledge support from the Sino-German Center for Research Promotion (No. M-0174). \\[3mm]

%





\bibliography{Spheremeasurement_GeesmannEtAl.bib}

\section*{End Matter}
\section{Preparation of the coaxial-thermocouple temperature sensors with the sphere}
The coaxial-thermocouple temperature sensors of the NSThM were fabricated as described in Wischnath {\itshape et al.}~\cite{Wischnath2008}. A platinum wire was threaded through a glass capillary which was then heated by an IR laser, thinned and ruptured. That way a tip is formed with a thin platinum wire, a few hundreds of nanometers in diameter, surrounded by a glass mantle. At the foremost end, the bare platinum wire is protruding a few tens of nanometers forming a sharp tip. The glass capillary was then fixed to a tip holder and a gold layer was evaporated onto it to form the second leg of the Pt-Au thermocouple. Usually, this type of tip is used to measure the extreme near-field effect of a laterally resolved variation of the heat transfer. For the measurements reported here, a 20~\textmu m silica sphere was glued with epoxy resin to the foremost end. Thereafter, a second gold film was evaporated to cover the silica sphere entirely by rotating the tip during the evaporation process. This process creates a conducting film from the backside of the tip to the foremost end of the tip allowing the application of a tunneling voltage between the sphere and the sample surface. Hence, contact of the sphere with the sample surface can be detected on an atomic level. 

\section{Theory}
To compare the experimental data with the theoretical prediction, a well-established technique is to determine the thermal power transferred between the gold sample and the gold-coated silica sphere by means of the proximity approximation~\cite{NatureEmmanuel}. It is assumed that the sphere of radius $R$ can be divided into a large number of concentric circular rings parallel to the planar sample surface. The transferred power $P_{\rm th}$ is given in good approximation by the mean heat flux $\phi = \langle S \rangle$ between two planar surfaces times the area of the circular ring, where $\langle S \rangle$ is the ensemble average of the Poynting vector component perpendicular to the sample surface. In the limit of an infinite number of circular rings of width $\rd r$, the ring area is $2 \pi r \rd r$ and summing up the contributions of all rings leads to the expression~\cite{NatureEmmanuel}
\begin{equation}
	P_{\rm th}(z) = \int_0^R \rd r\, 2 \pi r \phi(d(r)), 
	\label{Eq:PA}
\end{equation}
where $R$ is the radius of the sphere, $d(r) = z + R - \sqrt{R^2 - r^2}$ is the effective distance between each circular ring and the planar sample surface, and $z$ is the smallest distance between the sphere and the planar sample surface. The mean Poynting vector of the radiative heat flux between two planar gold samples is calculated using the standard expression of fluctuational electrodynamics.~\cite{PvH} The specific features of the surfaces enter with the permittivity of gold expressed by the Drude model $\epsilon_{\rm Au} (\omega) = 1 - \omega_p^2/[\omega (\omega + \ri \gamma)]$ with the plasma frequency $\omega_p = 1.4\times10^{16}~{\rm rad}/{\rm s}$ and the relaxation time $\gamma^{-1} = 3\times10^{-14}~{\rm s}$. 

The theoretical results from Eq.~(\ref{Eq:PA}) are compared to the experimental results. For this comparison, the expression
\begin{equation}
	P_{\rm ex}(z + z_0) = \frac{V_{\rm th}(z) + V_0}{s}
	\label{Eq:Ex}
\end{equation}
is used to fit the experimental data. Here, $V_{\rm th}(z)$ is the measured thermoelectric voltage. $V_0$ is a parameter to take the zero bias offset into account, which reflects the input offset voltage of the preamplifier. $s$ is a conversion factor with units V/W. The zero-distance correction $z_0$ accounts for surface roughness of the sphere, whereas the sample surface is assumed to be atomically flat. It corresponds to the non-zero distance between an ideal sample and sphere when any protrusion of the real sensor touches the gold sample, i.e. a tunneling current is detected. {Thus, the quantity $z+z_0$ represents the minimal separation between the surface of an ideal sphere and an absolutely flat sample surface (sphere-plate separation).} In the sphere-plate experiment conducted by Rousseau {\itshape et al.}~\cite{NatureEmmanuel} this zero distance correction had a value of about 150~nm for the measurement with a 22 \textmu m sphere due to some particles on the used sphere. The three parameters $V_0$, $s$, and $z_0$ are used to fit the experimental results $P_{\rm ex}$ to the theoretical value $P_{\rm th}$ obtaining the curves shown in Fig.~\ref{Fig:Data}.

\end{document}